\documentclass[letterpaper,twocolumn,10pt]{article}
\usepackage{usenix-2020-09}

\usepackage{tikz}
\usepackage{amsmath}
\usepackage{amssymb}
\usepackage{amsfonts}
\usepackage{booktabs}
\usepackage{balance}

\newcommand{\sys}{Skyplane}
\newcommand{\secref}[1]{\S{}\ref{#1}}
\newcommand{\figref}[1]{Fig.~\ref{#1}}

\usepackage{tabularx,colortbl}
\usepackage[most]{tcolorbox}
\definecolor{Gray}{HTML}{ABD5FF}  %

\usepackage[english]{babel}
\usepackage{soul}
\usepackage[short,nocomma]{optidef}
\usepackage{wrapfig}
\usepackage{subfig}

\newcommand{\allnotes}[1]{}

\newif\ifconfidential
\confidentialfalse  %
\ifconfidential
    \usepackage{fancyhdr}
    \usepackage{tcolorbox}
    \tcbuselibrary{breakable,skins,raster,listings}
    \tcbuselibrary{external}
    \renewcommand{\headheight}{16pt}
    
\fi

\begin{document}
\date{}
\title{\resizebox{\textwidth}{!}{\Large \bf \sys{}: Optimizing Transfer Cost and Throughput Using Cloud-Aware Overlays}}
\author{
{\rm Paras Jain, Sam Kumar, Sarah Wooders, Shishir G. Patil, Joseph E. Gonzalez, and Ion Stoica}\\
University of California, Berkeley
}

\maketitle

\begin{abstract}
Cloud applications are increasingly distributing data across multiple regions and cloud providers.
Unfortunately, wide-area bulk data transfers are often slow, bottlenecking applications.
We demonstrate that it is possible to significantly improve inter-region cloud bulk transfer throughput by adapting network overlays to the cloud setting---that is, by routing data through indirect paths at the application layer.
However, directly applying network overlays in this setting can result in unacceptable increases in cloud egress prices.
We present \sys{}, a system for bulk data transfer between cloud object stores that uses cloud-aware network overlays to optimally navigate the trade-off between price and performance.
\sys{}'s planner uses mixed-integer linear programming to determine the optimal overlay path and resource allocation for data transfer, subject to user-provided constraints on price or performance.
\sys{} outperforms public cloud transfer services by up to 4.6$\times$ for transfers within one cloud and by up to 5.0$\times$ across clouds.
\end{abstract}

\section{Introduction}
\label{sec:intro}

Increasingly, cloud applications transfer data across datacenter boundaries, both across multiple regions within a cloud provider (multi-region) and across multiple cloud providers (multi-cloud).
This is in part due to privacy regulations, the availability of specialized hardware, and the desire to prevent vendor lock-in.
In a recent survey~\cite{forrester}, more than $86\%$ of 727 respondents had adopted a multi-cloud strategy across diverse workloads.
Thus, support for fast, cross-cloud bulk transfers is increasingly important.

Applications transfer data between datacenters for batch processing (e.g. ETL~\cite{lakehouse}, Geo-Distributed Analytics~\cite{iridium}), and production serving (e.g. search indices~\cite{swan}).
Extensive prior work optimizes the throughput of bulk data transfers between datacenters within application-defined minimum performance constraints~\cite{amoeba, swan, b4, calendaring}.
All major clouds offer services for bulk transfers such as AWS DataSync~\cite{datasync}, Azure AzCopy~\cite{azcopy}, and GCP Storage Transfer Service~\cite{gcp_transfer_service}.

From the perspective of a cloud customer, transfer throughput and cost (price) are the two important metrics of transfers in the cloud.
Thus we ask \emph{how can we optimize network cost and throughput for cloud bulk transfers?}
We study this question in the context of designing and implementing \sys{}, an open-source cloud object transfer system.

A seemingly natural approach is to optimize the routing protocols in cloud providers internal networks to support higher-throughput data transfers.
Unfortunately, this is not feasible for two reasons.
First, rearchitecting the IP layer routing protocol to optimize for high-throughput bulk transfer could be negatively impact other applications that are sensitive to network latency.
Second, cloud providers lack a strong incentive to optimize data transfer to other clouds.
Indeed, AWS DataSync~\cite{datasync}, AzCopy~\cite{azcopy}, GCP Storage Transfer~\cite{gcp_transfer_service}, AWS Snowball~\cite{snowball}, and Azure Data Box Disk~\cite{azure_data_box}, all support data transfer \emph{into}, but not out of, their respective clouds.
Improvements to cross-cloud peering must be achieved with the cooperation of both the source and destination providers.

\begin{figure}[t]
    \centering
    \includegraphics[width=\linewidth]{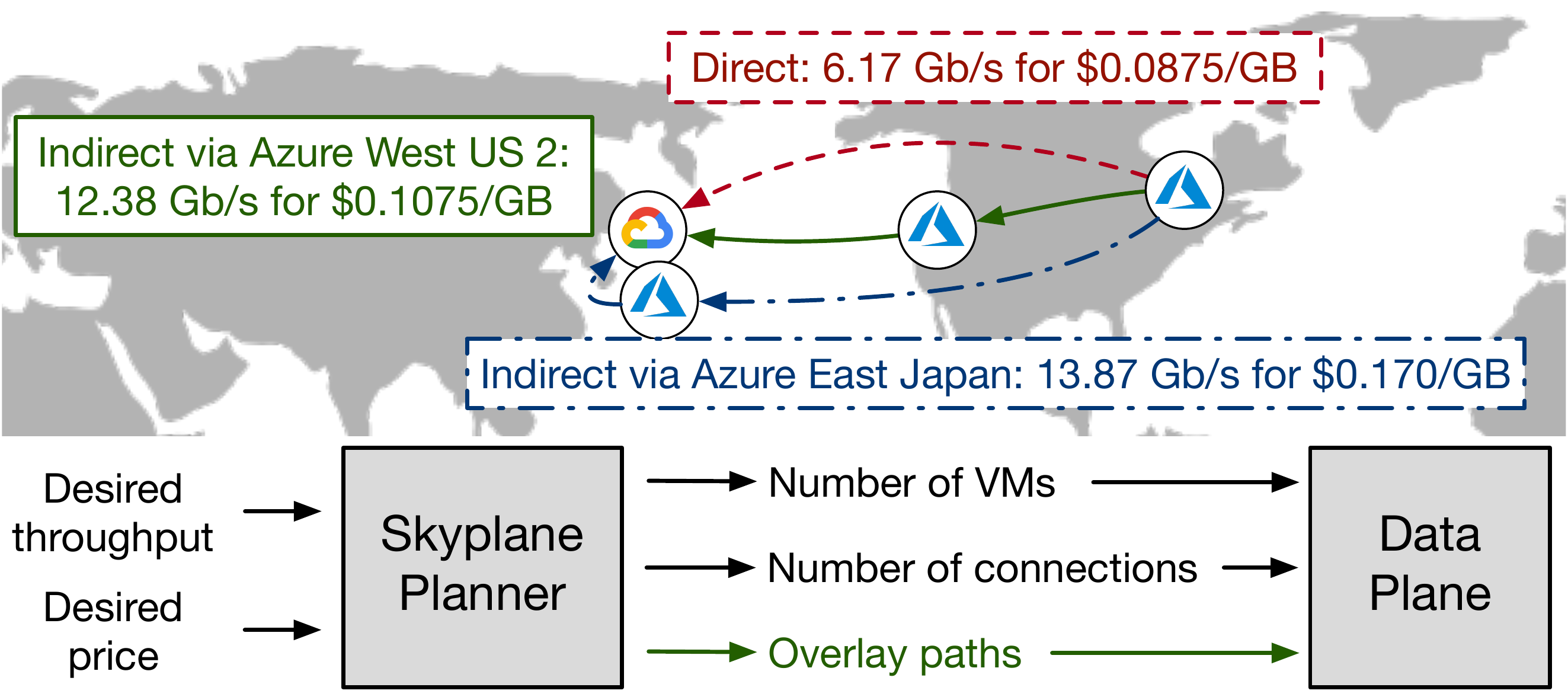}
    \caption{\textbf{Cloud-aware overlays:} \sys{} optimally transfers across cloud regions and providers subject to the user's cost and throughput requirements. \sys{} finds the visualized overlay path from Azure's \texttt{Central Canada} region to GCP's \texttt{asia-northeast1}, which is $2.0\times$ faster but just $1.2\times$ higher in price than the direct path.}
    \label{fig:teaser}
\end{figure}

\sys{}'s key observation is that we can instead identify \emph{overlay paths}---paths that send data via intermediate regions---that are faster than the direct path.
The throughput of the direct path from Azure's \texttt{Central Canada} region to GCP's \texttt{asia-northeast1} region is 6.2 Gbps.
Instead, \sys{} can route the transfer via an intermediate stop at Azure's \texttt{US West 2} with a throughput of 12.4 Gbps for a $2.0\times$ speedup (\figref{fig:teaser}).
Crucially, this can be implemented on top of the cloud providers' services without their explicit buy-in.

We are not the first to propose the use of overlay networks on the public Internet~\cite{ron}.
However, these techniques ignore two key considerations of public clouds: \textbf{price} and \textbf{elasticity}.

First, the highest-bandwidth overlay path may have an unacceptably high \textbf{price}.
Cloud providers charge for data egress separately for each hop along the overlay path.
To reduce the cost of the overlay, it is essential to transfer data along cheap paths to trade off price and performance.
For example, in \figref{fig:teaser}, one can achieve 13.9 Gbps by instead using Azure's \texttt{East Japan} region as the relay, but the cost would be $1.9\times$ that of transferring data directly.
In contrast, using Azure's \texttt{West US 2} region has only a $1.2\times$ cost overhead with similar performance.
Thus, \sys{} operates in a richer \textit{problem space} than traditional application-level routing---one where cloud instance and cloud egress fees are significant.

Second, whereas the bandwidth between two nodes in a traditional network overlay~\cite{ron} is considered ``fixed,'' in \sys{}'s setting it depends on \textbf{elasticity}---the ability to allocate more resources at each cloud region.
For example, one can increase the capacity of any overlay path by simply allocating more VM instances in each cloud region.
There are a limited number of physical machines at each cloud region, which cloud providers pass on to users in the form of instance limits.
An overlay enables improved throughput beyond this limit.
Thus, \sys{} operates in a richer \textit{solution space} than traditional application-level routing---one where we must choose the number of VMs to use as relays due to cloud elasticity.

\sys{} addresses both \textbf{price} and \textbf{elasticity}, empowering users to navigate the trade-off between price and performance while leveraging the elasticity of cloud resources.
Users can ask \sys{} to maximize bandwidth subject to a cost ceiling, or minimize cost subject to a bandwidth floor.

At the heart of \sys{} is a planner that computes a data transfer plan, subject to the user's constraints, that specifies the overlay path to use and amount of cloud resources to allocate along that path.
Price and elasticity make it challenging to compute the plan.
Our insight is that, with some care, planning can be formulated as \emph{linear} constraints.
Thus, \sys{}'s planner can discover the optimal plan by solving a mixed-integer linear program (MILP), or closely approximate the optimal plan by solving a relaxed linear program (LP).
Both can be accomplished using a fast, off-the-shelf solver.

Our \sys{} prototype\footnote{\url{https://github.com/skyplane-project/skyplane}}outperforms AWS DataSync by up to 4.6$\times$ and GCP Storage Transfer by up to 5.0$\times$.
\sys{} also outperforms academic baselines such as RON by 34\% while reducing cost by 62\%.

\section{Background}

\paragraph{Network overlays}
In the early 2000s, network overlays emerged as a technique for \textit{application-level routing} without the \textit{participation of underlying network providers}.
These network overlays can be designed to improve performance or reliability.
Notable network overlays include Chord~\cite{chord}, Resilient Overlay Networks (RON)~\cite{ron}, Bullet~\cite{bullet}, Baidu BDS~\cite{bds} and Akamai's backbone~\cite{akamai,akamai2_sosp}.

Although ISPs may have broad visibility into their networks, the metrics that ISPs use to select routes may not align with user preferences.
Wide-area networks today do not allow specification of alternative routing preferences while network overlays provide applications a mechanism to control routing decisions.
For example, Akamai uses a network overlay to reduce the latency of CDN misses while RON routes around network outages via an unaffected intermediate host.

RON is implemented by periodically measuring network performance via probes embedded in a fixed set of routers.
When path outages occur, RON selects an intermediate relay router to circumvent the outage.
This intermediate router is selected to have low packet loss or latency to/from the client and server.
Optionally, RON can use a model of TCP Reno's throughput~\cite{tcp_equation} to select intermediate routers.
RON will generally select only a single intermediate node.

\vspace{-5pt}
\paragraph{Wide-area networking in the cloud}\label{s:background_wan_cloud}
From the perspective of cloud customers, the cloud is \emph{elastic}---additional resources can be allocated on demand.
For example, an overloaded cloud application can leverage the cloud's elasticity by allocating additional VM instances.
However, the physical reality of the cloud is that there are only finite resources at each region.
Therefore, cloud providers impose \emph{service limits} on their customers for resources such as VMs.

Each VM's network bandwidth is throttled according to its instance type.
For example, an AWS \texttt{m5.8xlarge} instance can use at most 10 Gbps of network bandwidth, and an Azure \texttt{Standard\_D32\_v5} instance can use at most 16 Gbps of network bandwidth.
Furthermore, only some of the available bandwidth can be used for egress traffic to another cloud provider.
The policies differ by cloud provider.
AWS limits VM egress bandwidth to the larger of 5 Gbps or 50\% of total bandwidth~\cite{aws_vm_egress_limit}, GCP limits VM egress bandwidth to any public IP address to 7 Gbps~\cite{gcp_vm_egress_limit}, and Microsoft Azure has no egress limit beyond the total bandwidth limit for a VM.
Of course, the actual achievable TCP network bandwidth is subject to congestion control which may be less than the limit.

\vspace{-5pt}
\paragraph{Cloud egress pricing}
Cloud providers charge egress prices for network traffic leaving a cloud region.
Importantly, egress prices are assessed based on the \emph{volume} of data transferred, not the rate at which it is transferred.
Transferring a file at 10 Mbps or at 10 Gbps will result in the same egress charge.
Egress charges introduce asymmetry in billing---there is no corresponding ingress charge for transfers into a cloud.

For intra-cloud transfers (i.e., transfers between two regions or zones in the same cloud), transfers between geographically distant endpoints are priced more than transfers between nearby endpoints.
In contrast, inter-cloud transfers (i.e., transfers between two cloud providers) are billed at the same rate regardless of the transfer's geographic distance.
For example, the egress price from a single Azure region is billed at the same rate for \emph{any} destination outside of Azure, including any region in AWS or GCP~\cite{aws_bandwidth_pricing, azure_bandwidth_pricing, gcloud_network_pricing}.

Egress prices typically dominate the cost of a bulk transfers.
For example, if a VM sends data at a rate of 1 Gbps for an hour on AWS with an Internet egress price of \$0.09/GB, the total egress charge will total \$40.50, which far exceeds the VM price of \$1.50 (for \texttt{m5.8xlarge})~\cite{aws_bandwidth_pricing}.

\vspace{-5pt}
\paragraph{Cloud object storage}
AWS, Azure, and GCP provide object storage APIs that allow customers to save data attached to a string key.
Data is stored immutably and therefore any updates require writing a new version.
Unlike POSIX file systems, object stores do not provide atomic metadata operations (e.g., rename).
Consistency models vary across providers.
Cloud object stores store copies of a blob on multiple machines to improve availability and durability.
Large objects support concurrent writes via sharding.
Read throughput of a single shard may be limited by the provider (e.g. 60 MB/s for Azure~\cite{azurescalabilitytarget}).
\section{Overview of \sys{}}

\sys{} allows applications to efficiently transfer large objects from an object store in one region to an object store in another cloud region or provider.
To use \sys{}, the user installs the \sys{} client locally and configures it with access to cloud provider-supplied credentials.
Then, the user submits a job, together with a constraint on price or bandwidth.
The job specifies which objects to transfer, the source cloud provider and region, and the destination cloud provider and region.
The constraint can have one of two forms: it can ask \sys{} to optimize either \textit{bandwidth subject to a price ceiling}, or \textit{price subject to a bandwidth floor}.

\begin{figure}[t]
    \centering
    \includegraphics[width=\linewidth]{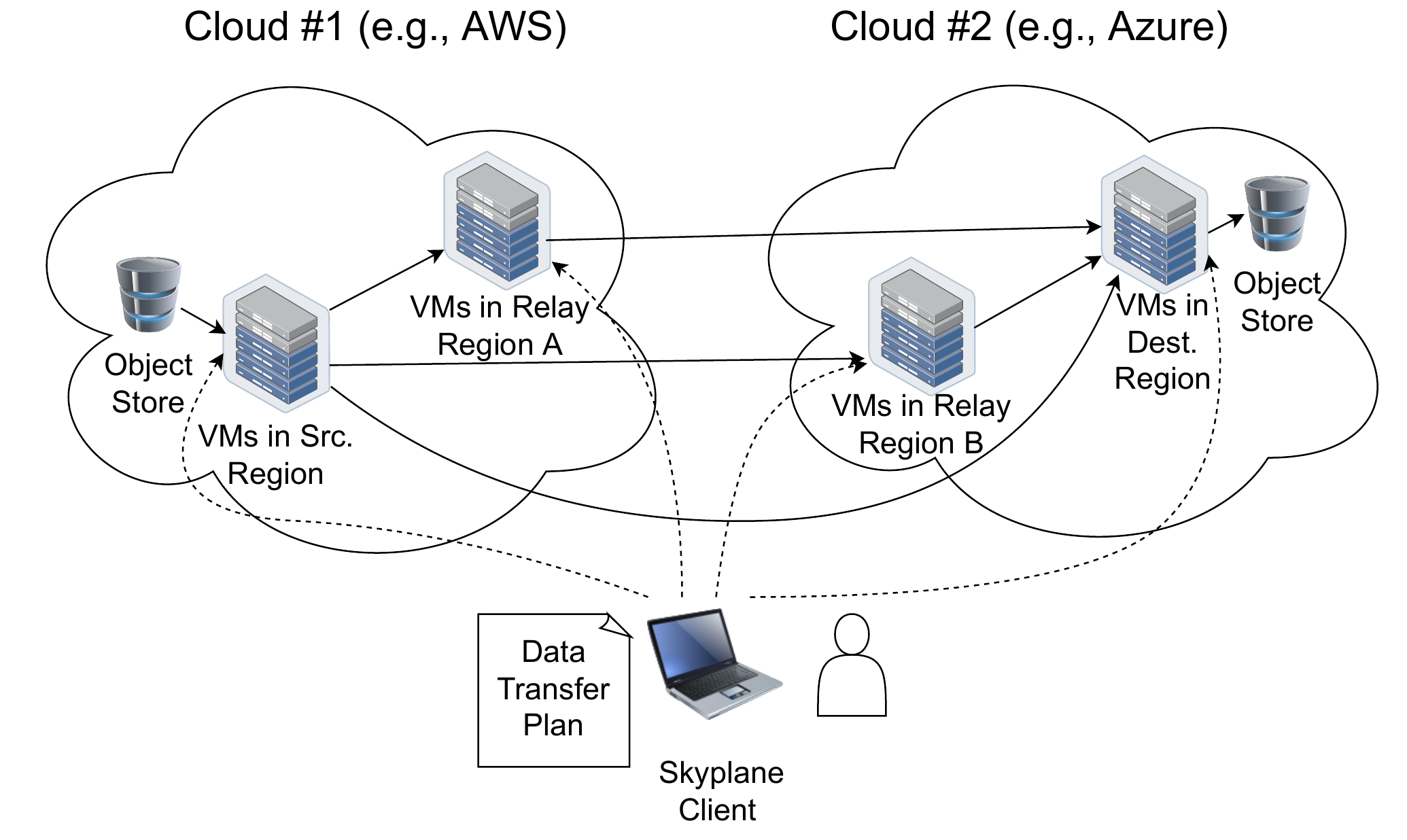}
    \caption{\sys{} splits an example data transfer over three paths: the direct path, and two indirect paths. Dashed lines indicate control orchestration (e.g., for spawning VMs) and solid lines depict the flow of object data.}
    \label{fig:data_plane}
\end{figure}

\sys{} itself comprises a \emph{planner}~(\figref{fig:teaser}, bottom) and a \emph{data plane}~(\figref{fig:data_plane}).
Given the user's job and constraint, the planner produces an optimal data transfer plan to complete the job subject to the constraint.
The planner relies on a profile of the network throughput between different cloud regions.
The data plane is responsible for executing the data transfer plan: allocating cloud resources (e.g., VMs), transferring data between them, and interacting with object stores.

\begin{figure*}[t]
    \centering
    \includegraphics[width=.7\linewidth]{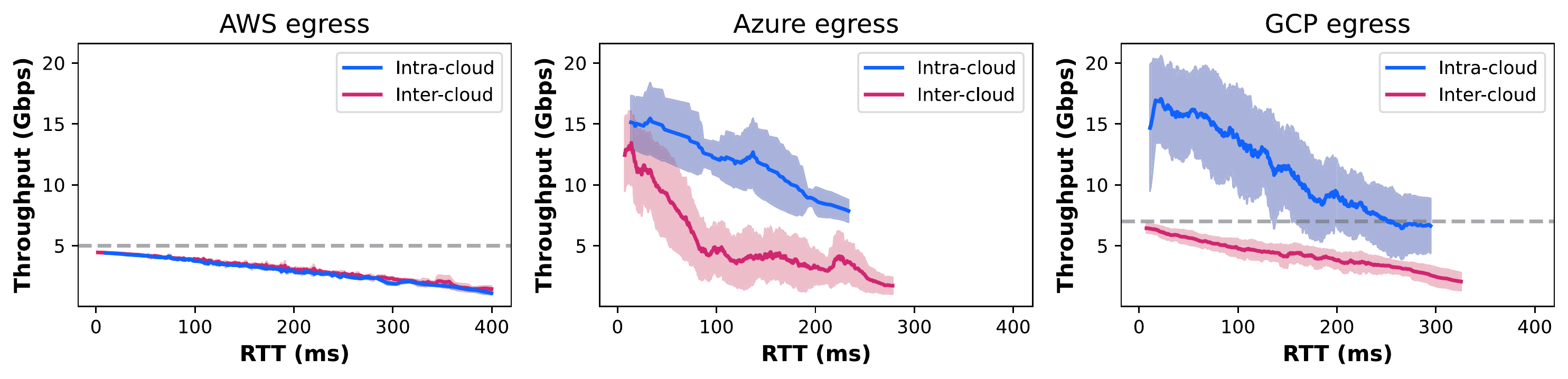}
    \caption{\textbf{Intra-cloud vs. inter-cloud links}: Inter-cloud links are consistently slower than intra-cloud links for network routes from Azure and GCP. Service limits are shown with a dashed line; GCP throttles inter-cloud egress to 7 Gbps while AWS throttles \textit{all egress traffic} to 5 Gbps.}
    \label{fig:profile:inter_vs_intra_cloud}
\end{figure*}

\begin{figure*}[t]
    \centering
    \includegraphics[width=0.7\linewidth]{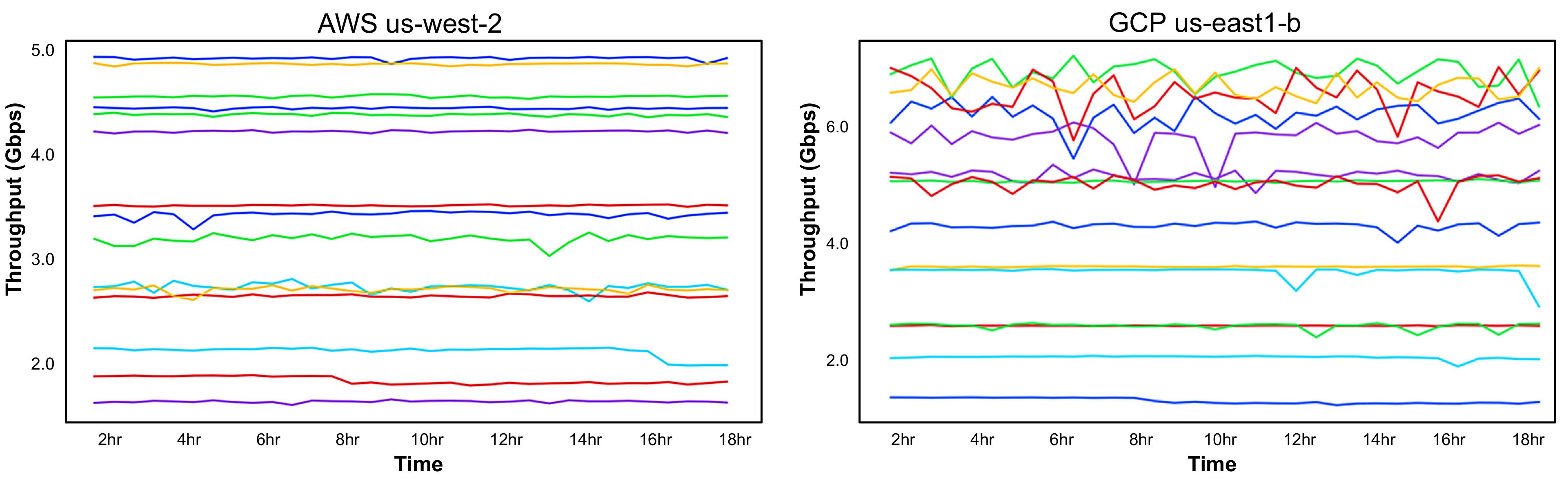}
    \caption{\textbf{Stability of egress flows over 18 hour period}: Continuous probes of cloud networks over one day reveal that routes from AWS have stable throughput over time. Paths between GCP regions are noisy but have a consistent mean.}
    \label{fig:profile:stability}
\end{figure*}

\subsection{\resizebox{0.425\textwidth}{!}{Overlay formulation in \sys{}'s planner}}
Suppose the user needs to transfer an object from a source cloud region, $A$, to a destination cloud region, $B$.
A na\"ive object transfer system might spawn VMs in regions $A$ and $B$, and transfer data via a TCP connection between the two VMs.
\sys{} improves performance compared to this baseline by applying principles from overlay networks~\cite{ron}.
For example, \sys{} may identify a third cloud region, $C$, and transfer data from $A$ to $B$ via $C$.
This is accomplished at the application layer; \sys{} will spawn a VM in region $C$, set up TCP connections from $A$ to $C$ and from $C$ to $B$.
We refer to intermediate regions like $C$ as \emph{relay regions}.

The baseline approach ($A\rightarrow B$) routes data along the ``direct path,'' since it uses the default path provided by the Internet.
However, \sys{} ($A\rightarrow C\rightarrow B$) routes data along the an ``indirect path,'' that may not be on the Internet-provided default path.
An indirect path may use multiple relays although a single relay is usually sufficient.

A key difference between \sys{} and classical overlay networks is that \sys{} takes price into account when choosing the overlay path to use for a job.
Concretely, \sys{}'s planner uses a \emph{price grid} and a \emph{throughput grid} to determine which indirect path to use.
The price grid specifies the price of transferring data between each pair of cloud regions, in each direction.
We computed the price grid based on information on the cloud providers' websites and from querying the cloud APIs.
The throughput grid is collected by measuring the network, as we explain in the next subsection.

Note that throughput grid measurements are made using TCP connections, subject to TCP congestion control.
Thus, the throughput grid measures the bandwidth available to \emph{a single user} for transferring data, accounting for cross-traffic from other users' flows.
We assume a high degree of statistical multiplexing in wide-area network traffic---in other words, that the bandwidth consumed by a single user's bulk transfer is negligible compared to the total available inter-region bandwidth.
This allows a \sys{} user to compute a data transfer plan without regard to other users' bulk transfers using \sys{} or other bulk transfer tools---all cross traffic from other users is assumed to be accounted for in the throughput grid.
As we show in the next subsection, the bandwidth of inter-region TCP connections is relatively stable in the short term, validating our assumption of high statistical multiplexing.

\subsection{Profiling cloud networks}
The planner relies on a profile of the network throughput between pairs of cloud regions.
We collected a throughput grid by measuring the TCP goodput between each region pair using \texttt{iperf3}.
In total, computing this profile cost approximately \$4000 in egress charges.

\figref{fig:profile:inter_vs_intra_cloud} displays the relationship between network latency and throughput for profiling routes originating from GCP and Azure for our measured throughput grid.
For GCP, we leverage internal IPs which improve intra-cloud bandwidth.
For both GCP and Azure, intra-cloud routes had lower tail RTTs than inter-cloud routes.
We observe that in both GCP and Azure, inter-cloud links are slower than intra-cloud links.
As Azure has no service limit for egress bandwidth, we see the fastest intra-cloud links achieve up to the NIC capacity of 16 Gbps.
However, both GCP and AWS encounter egress throttling at 7 Gbps and 5 Gbps respectively.

A natural question is how frequently the throughput grid must be re-measured.
\figref{fig:profile:stability} visualizes achieved throughput from AWS \texttt{us-west-2} and GCP \texttt{us-east1-b} taken every 30 minutes over an 18 hour timespan.
Throughput is very stable over time for both inter-cloud and intra-cloud routes from AWS \texttt{us-west-2}.
Routes from GCP \texttt{us-east1-b} to AWS destinations is similarly very stable but intra-cloud routes to GCP destinations are less stable.
Regardless, the overall rank order of regions by throughput remains mostly consistent over medium-term timescales.
Thus, it should be sufficient to profile networks relatively infrequently (i.e. every few days).
In practice, this information could be collected by third-party service, or measured via active probing along live transfers.

\subsection{\sys{}'s data plane}
\sys{}'s data plane executes data transfers using the plan computed by \sys{}'s planner.
Ephemeral VMs for a single transfer, called ``gateways,'' are provisioned in the source region, destination region, and overlay regions for a transfer plan.
Each source gateway reads a small shard of data from the object store and transfers data via intermediate gateways to the destination where the shard is written.

\sys{} reads data from an object store in the source cloud region and writes data to an object store in the destination cloud region.
We focus on the object stores provided as a service by AWS S3, Azure Blob Storage, and Google Storage.
Unlike a traditional overlay network, there is no central \sys{} service that allocates resources to each user from a pool of ``\sys{} resources.''
Instead, \sys{} can be understood as a local service run by each user that is invoked when an application needs to transfer data.
\sys{} directly allocates cloud resources on the user's behalf when processing a job, and manages those resources to transfer the user's data across cloud regions.
This allows \sys{} to allocate and manage each user's resources according to their cost and performance objectives, independently from the cloud providers' existing data transfer services, while relying on clouds to offer a large pool of resources and manage isolation between users.

\section{Principles of \sys{}'s planner}\label{sec:planner}
\begin{figure}[t]
    \centering
    \includegraphics[width=0.8\linewidth]{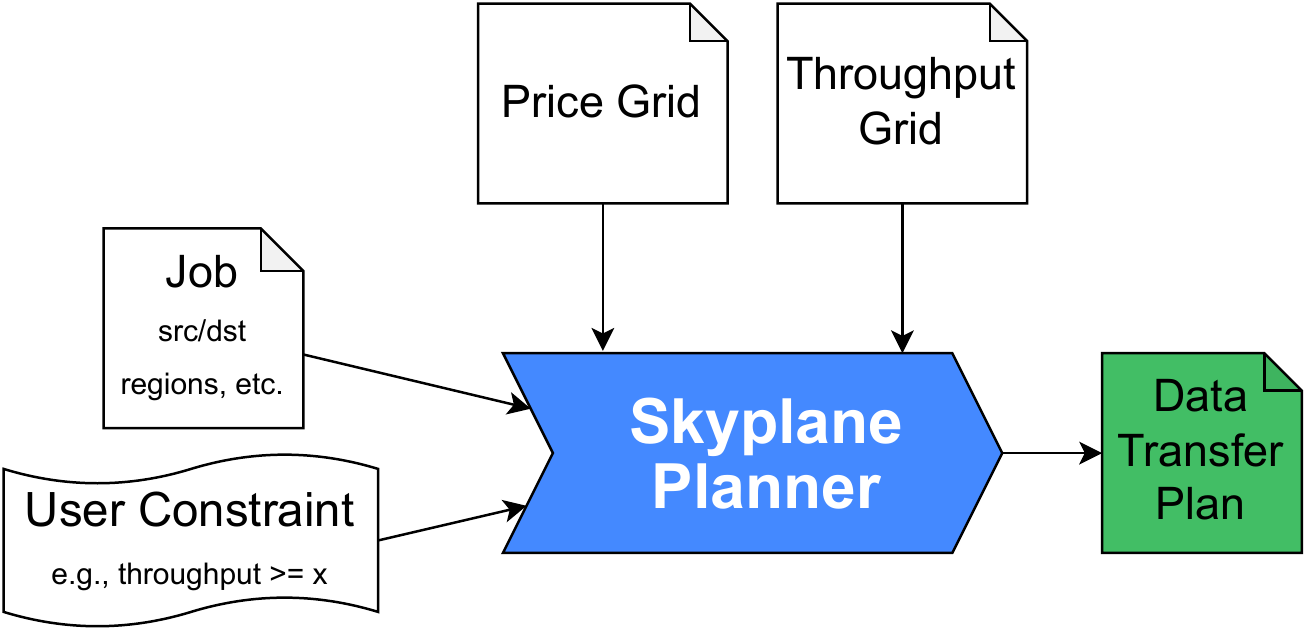}
    \caption{\sys{}'s planner considers throughput and cost constraints from the user along with per-cloud price information and an inter-region throughput profile grid to determine the optimal data transfer plan.}
    \label{fig:planner_flowchart}
\end{figure}

\sys{}'s planner\footnote{Explore \sys{}'s planner at \url{https://optimizer.skyplane.org}} is responsible for developing a plan for transferring data across the wide area to complete an object transfer job submitted by a user or their application~(\figref{fig:planner_flowchart}).
This plan describes the overlay path and the amount of cloud resources to allocate along that path to facilitate the transfer.

\vspace{0.2em}
\sys{}'s planner supports two modes:

\vspace{0.2em}
\noindent\textbf{Cost minimizing}: The planner will minimize cost subject to an application-specified throughput constraint.

\vspace{0.2em}
\noindent\textbf{Throughput maximizing}: The planner will maximize throughput subject to an application-specified cost constraint.
\vspace{0.2em}

As we will describe in \secref{sec:solver}, \sys{} finds the optimal plan by formulating it as an Mixed-Integer Linear Program (MILP) and using a fast but exponential-time solver.
This section describes the degrees of freedom available to the optimizer to navigate the price-performance trade-off for the user's specified constraint.
Our goal is to describe what aspects of the plan are at the planner's disposal, justify why it is reasonable to vary those aspects of the plan, and describe certain techniques available to the planner to manage the price-performance trade-off.
Note that the planner is not directly programmed to use these techniques; they are merely patterns that it discovers in the course of finding the optimal MILP solution.

\subsection{Achieving low instance and egress costs}\label{s:price}

That bandwidth costs dominate the cost of data transfer (\secref{s:background_wan_cloud}) is both a challenge and an opportunity for \sys{}.
It is an opportunity because it allows \sys{} to be competitive with the price of using data transfer tools provided directly by the cloud providers (e.g. AWS DataSync, AzCopy, GCP Cloud Transfer Service), as those tools incur bandwidth costs but not instance costs.
It is a challenge for \sys{} because it implies that, used na\"ively, indirect paths are much more expensive than direct paths.
This is because egress bandwidth is charged for each hop along the path.
For example, for a path $A\rightarrow C\rightarrow B$, the bandwidth cost must be paid for both $A\rightarrow C$ and $C\rightarrow B$, which could be \emph{double} the cost of transferring over the direct path.
As a result, it is crucial for \sys{}'s optimizer carefully manage egress transfer costs.

\subsubsection{Choosing the relay region}

One way for \sys{} to manage the additional cost associated with indirect paths is to carefully choose the relay region $C$ to minimize this cost.
For example, suppose that a user needs to transfer an object from AWS \texttt{us-west-2} (region $A$) to Azure \texttt{UK South} (region $B$).
The direct path $A\rightarrow B$ would require the user to pay \$0.09 per GB, the cost of bandwidth leaving AWS' network.
If the relay region $C$ is chosen in \texttt{us-central-1} or \texttt{us-east-1}, then the overall bandwidth price will only increase slightly; while the $C\rightarrow B$ transfer still incurs \$0.09 per GB, as data is leaving AWS' network, the $A\rightarrow C$ bandwidth only costs \$0.02 per GB, as it is an intra-continental transfer within the cloud provider's network.
\sys{}'s planner can use the throughput and price grids to identify relay regions that improve the performance of the transfer while minimizing additional bandwidth costs.

\subsubsection{Combining multiple paths}

Another way to manage the cost of indirect paths is to split the data transfer over multiple paths, in order to make fine-grained trade-offs between price and performance.
For example, suppose that \sys{} identifies a high-bandwidth indirect path, but that the path is more expensive than the user's price ceiling.
\sys{} can still benefit partially from that indirect path by sending part of the data over that path, at higher cost, and the remaining data over the direct path $A\rightarrow B$, at lower cost.
Thus, \sys{} may average the price and performance of multiple paths, when doing so allows \sys{} to more optimally satisfy the user's constraints.

\subsection{Parallel TCP for high bandwidth}\label{s:parallel_tcp}

\sys{} uses parallel TCP connections---that is, bundles of TCP connections---to achieve high goodput over a chosen path.
This is a well-known technique for achieving good performance, particularly for wide-area transfers~\cite{gridftp, psockets}.
Our \sys{} implementation uses up to 64 outgoing connections for each VM instance, as we empirically measured that using additional connections typically resulted in diminishing benefits in aggregate goodput.
When collecting measurements for the throughput grid, we make sure to use 64 parallel connections to measure the achievable TCP goodput for each ordered pair of regions.

It is known that using multiple TCP streams in parallel may cause an application to obtain more than its ``fair share'' of bandwidth~\cite[\S A.1]{floydcongestion}, particularly in contexts where networks are running at nearly 100\% utilization~\cite{b4}.
Our view is that, despite this, it is acceptable to use multiple TCP connections in parallel in the context of \sys{}.
There are three reasons for this.
First, it is common for applications to use parallel TCP, including for workloads like bulk data transfer~\cite{gridftp, computernetworking}.
It is important for \sys{} to appropriately compete with such applications for limited bandwidth.
Second, the user pays the cloud provider for bandwidth, both in the form of the bandwidth price (total amount transferred) and the instance price (rate at which data can be transferred), and it is natural for users to be able to make use of the bandwidth that they pay for.
Third, cloud providers control the datacenter network, and can shape traffic in the presence of congestion to ensure that each customer gets a fair share of bandwidth.

\subsection{Multiple VMs for high bandwidth}\label{s:multiple_instances}

For a given overlay path, \sys{} must allocate sufficient resources along the path to achieve high bandwidth.
However, the achievable outgoing bandwidth from a VM instance is limited, as described in \secref{s:background_wan_cloud}.

Therefore, \sys{} may allocate multiple VM instances at certain regions along the path, to increase \emph{aggregate} data transfer rate of the VMs at each region.
Although simply using larger VMs may seem like a viable alternative, it is less effective than using multiple instances due to per-instance bandwidth limits.
\sys{} uses a fixed VM size, and its planner chooses how many instances to allocate in each region, under the assumption that TCP goodput scales linearly with the number of allocated VM sizes.

It may seem that \sys{} can achieve an arbitrarily high bandwidth by spawning many instances in each region.
Unfortunately, this simple strategy does not work because cloud resources are not perfectly elastic.
The finite capacity for VMs in a datacenter is passed down to cloud customers in the form of service limits, which limit the number of VM instances, and therefore the amount of network bandwidth, that users can allocate in each region.
While users can request limit increases, these are ultimately subject to resource availability.
To model this, \sys{}'s planner takes into account a limit on the number of instances that a user can allocate per region.

\section{Finding optimal transfer plans}\label{sec:solver}
\sys{}'s planner searches for cost-efficient high-throughput transfer plans that jointly specify the overlay path, TCP connections between regions and VMs to provision per region.

At the core of \sys{}'s planner is an optimizer that finds the optimal plan using off-the-shelf Linear Programming (LP) solvers. We formalize the constraints of our problem as Mixed Integer LP (MILP) which can quickly be solved in under 5 seconds with an open-source solver. The problem can be further relaxed into a continuous LP which is solvable in worst-case polynomial time via interior point methods~\cite{karmarkar_poly_time_lp}.

Independently optimizing for each variable then combining partial solutions would not guarantee a globally optimal solution. It is therefore important that \sys{}'s planner models all variables in an integrated search space to obtain provably optimal data transfer plans.

\newcommand{\flow}[0]{\textsc{F}}
\newcommand{\instances}[0]{\textsc{N}}
\newcommand{\connections}[0]{\textsc{M}}

\newcommand{\flowmax}[0]{\small{\textsc{Limit}}^{link}}
\newcommand{\tmin}[0]{\textsc{tput goal}}
\newcommand{\volume}[0]{\textsc{volume}}
\newcommand{\connectionlimit}[0]{\small{\textsc{Limit}}^{conn}}
\newcommand{\frob}[2]{\langle #1, #2 \rangle}
\newcommand{\maxinstances}[0]{\small{\textsc{Limit}}^{VM}}
\newcommand{\servicelimiti}[0]{\small{\textsc{Limit}}^{ingress}}
\newcommand{\servicelimite}[0]{\small{\textsc{Limit}}^{egress}}
\newcommand{\coste}[0]{\textsc{Cost}^\textit{egress}}
\newcommand{\costi}[0]{\textsc{Cost}^\textit{VM}}

\begin{table}[t]
\centering
\resizebox{0.8\linewidth}{!}{
\begin{tabular}{ll}
\toprule
\multicolumn{2}{c}{\textbf{Variables}} \\ 
$\flow \in \mathbb{R}_+^{|V| \times |V|}$ & \textit{Throughput grid} \\ 
$\instances \in \mathbb{Z}_+^{|V|}$ & \textit{VMs per region} \\ 
$\connections \in \mathbb{Z}_+^{|V| \times |V|}$ & \textit{TCP conn. per region} \\ 
\multicolumn{2}{c}{\textbf{Constraint: goal throughput}} \\ 
$\tmin \in \mathbb{R}_+^{|V| \times |V|}$ & \textit{User's desired throughput} \\ 
\multicolumn{2}{c}{\textbf{Constants: provider limit}} \\ 
$\flowmax \in \mathbb{R}_+^{|V| \times |V|}$ & \textit{Throughput grid limit} \\ 
$\connectionlimit \in \mathbb{Z}_+^{|V| \times |V|}$ & \textit{TCP connection limit} \\ 
$\servicelimiti \in \mathbb{Z}_+^{|V|}$ & \textit{VM limit} \\ 
$\servicelimite \in \mathbb{Z}_+^{|V|}$ & \textit{Egress bandwidth limit} \\ 
\multicolumn{2}{c}{\textbf{Constants: provider cost}} \\ 
$\coste \in \mathbb{R}_+^{|V|}$ & \textit{Egress cost (\$/Gbit)} \\ 
$\costi \in \mathbb{R}_+^{|V|}$ & \textit{VM cost (\$/s)} \\ \bottomrule
\end{tabular}
}
\caption{Symbol table for \sys{}'s ILP formulation.}
\label{tab:symbols}
\end{table}

\subsection{Cost minimizing overlay paths}
Flow networks can naturally represent overlay networking topologies like those used by Akamai~\cite{akamai}. We start with a min-cost flow problem. The following primal LP finds the optimal flow matrix $\flow \in \mathbb{R}_+^{|V| \times |V|}$ for a network topology graph $G = (V, E)$ where nodes represent regions and edges are links:

\begin{argmini}|l|
{\flow}{\frob{C}{\flow}}{}{}
\addConstraint{\textstyle\sum_{(c, v) \in E} \flow_{c, v}}{\geq \tmin}{}
\addConstraint{\textstyle\sum_{(u, v) \in E} \flow_{u, v}}{= \textstyle\sum_{v, w} \flow_{v, w}}{~~~\forall v \in V - \{s, t\}}
\addConstraint{0 \leq \flow}{\leq \flowmax}{}
\end{argmini}
where $s$ and $t$ are the source and destination regions, $\flowmax \in \mathbb{R}_+^{|V| \times |V|}$ is the maximum capacity for each link and ${C \in \mathbb{R}_+^{|V| \times |V|}}$ is the cost per unit of bandwidth between regions. We use the same notation for matrix and vector inner products: $\frob{C}{\flow} = \sum_{u,v} {C}_{u,v} {\flow}_{u,v}$.

\subsubsection{Objective: Minimize cost from egress and VMs}
Min-cost flows do not accurately reflect the cost of transfers in the cloud. The total cost of a transfer in \sys{} includes \textit{egress cost} and \textit{VM cost}. Note that this objective is not linear; we present a linear reformulation in Sec.~\ref{sec:solver:linearizing_objective}. We present the full objective is in the in Equation~\ref{formal:full_objective}.

\paragraph{Modeling egress cost}
Unlike physical networks, virtual networks in the cloud will charge the same amount if 1GB of data is sent at 1 Mbps or 10 Gbps. Transfers are priced according to \textit{egress volume} (\$ per GB, $\coste$) rather than \textit{bandwidth} (\$ per Gbps). We can update the cost function to instead model the transfer cost by first computing how much the overlay path costs to run per unit time and then scale that by the runtime for a transfer. 
We denote the total volume of the transfer as $\volume$. Total egress cost is then:
\begin{equation}
\label{eqn:solver:egress_objective}
\underbrace{\frob{\flow}{\coste}}_{\text{Egress cost per s}} * \underbrace{\volume \div \textstyle\sum_{v \in V} \flow_{s, v}}_{\text{Transfer time}}
\end{equation}

\paragraph{Modeling VM cost}
Multiple VMs can increase aggregate bandwidth as discussed in Sec.~\ref{s:multiple_instances}. To optimally trade-off parallel VMs with the overlay, we introduce a new decision variable $\instances \in \mathbb{Z}_+^{|V|}$ that models the number of instances use to transfer data per region. VM count per region may vary due to asymmetric egress and ingress limits. To accurately consider transfer costs from VMs, we add the the following instance cost expression to Equation~\ref{eqn:solver:egress_objective} where $\costi$ is a vector containing the cost per second per VM in each region:
\begin{equation}
\label{eqn:solver:instance_objective}
\underbrace{\frob{\instances}{\costi}}_{\text{VM cost per s}} *
\underbrace{\volume \div \textstyle\sum_{v \in V} \flow_{s, v}}_{\text{Transfer time}}  %
\end{equation}

\paragraph{Linear reformulation of the objective}
\label{sec:solver:linearizing_objective}
As written, the objective in Equation~\ref{formal:full_objective} is not linear due to a product of variables between $\flow$ and $\instances$. By reformulating the problem to instead consider finding a plan that provides \textit{exactly} $\tmin$ (instead at least), the runtime for the transfer can be reduced to a constant $\volume \div \tmin$.

\subsubsection{Constraints: Cloud provider service limits}
Resources are not infinite at cloud regions; providers limit the number of VMs that a user may provision and in some cases, providers may throttle the performance of ingress and egress.

\paragraph{Per VM ingress and egress limits}
AWS and GCP each throttle egress from their clouds via SDN policies. For AWS, instances with 32 cores or less are limited to 5 Gbps. For GCP, individual flows are limited to 3 Gbps and total egress is service limited to 7 Gbps. Ingress is bottlenecked by VM NIC bandwidth. We constrain the maximum ingress bandwidth per VM to $\servicelimiti$ via Constraint~\ref{formal:instance_ingress} and the maximum egress bandwidth per VM to $\servicelimite$ via Constraint~\ref{formal:instance_egress}.

\paragraph{Constraining TCP connections}
Using parallel TCP connections is a well known approach to improve WAN performance as discussed in Section~\ref{s:parallel_tcp}. Yet, bandwidth does not scale linearly with connections (Figure~\ref{fig:eval:system_ablation:tcp_connections}). We introduce a decision variable $\connections \in \mathbb{Z}_+^{|V| \times |V|}$ representing the number of connections between a \textit{pair of regions} (not per VM pair). Constraint~\ref{formal:flow_cap} ensures $\connections$ is constrained by $\instances$ and $\connectionlimit$ (typically 64 per VM). We then limit the total incoming and outgoing connections with Constraints~\ref{formal:connection_ingress} and~\ref{formal:connection_egress}.

\paragraph{Per-region VM limits}
We introduce the variable $\instances \in \mathbb{Z}_+^{|V|}$ to denote the number of VMs per region. $\instances$ must be under the global instance cap in Constraint~\ref{formal:max_instancess}. The optimizer linearly scales the maximum number of egress TCP connections per region by the number of VMs provisioned in each region.

\subsubsection{Continuous relaxation of MILP}
To improve solve times, $\instances$ and $\connections$ are relaxed into real valued variables $\instances \in \mathbb{R}_+^{|V|}$ and $\connections \in \mathbb{R}_+^{|V| \times |V|}$. Rounding variables down performs comparably to randomized rounding with solutions $\le 1 \%$ from optimal. The relaxed problem has worst case polynomial time complexity~\cite{karmarkar_poly_time_lp}.

\subsubsection{Full formulation of the cost optimal solver}
All variables and constants are listed in Table~\ref{tab:symbols}. The full formulation of \sys{}'s optimizer is:
\begin{argmini!}|l|[1]
{\shortstack{\flow, \instances\\\connections}}{\frac{\volume}{\tmin}  \big(\frob{\flow}{\coste} ~+\frob{\instances}{\costi}\big)}{}{\label{formal:full_objective}}
\addConstraint{\flow \leq}{(\flowmax \odot \connections)\div \connectionlimit\label{formal:flow_cap}}{}
\addConstraint{\textstyle\sum_{v \in V} \flow_{s, v}}{\geq \tmin\label{formal:flow_conservation_source}}{}
\addConstraint{\textstyle\sum_{u \in V} \flow_{u, t}}{\geq \tmin\label{formal:flow_conservation_sink}}{}
\addConstraint{\textstyle\sum_{u \in V} \flow_{u, v}}{= \textstyle\sum_{u \in V} \flow_{v, u}~~~~~~\forall v \in V-\{\text{s}, \text{t}\}}{\label{formal:flow_conservation_other}}
\addConstraint{\textstyle\sum_{u \in V} \flow_{u, v}}{\leq \servicelimiti_v * \instances_v~~~\forall v \in V}{\label{formal:instance_ingress}}
\addConstraint{\textstyle\sum_{v \in V} \flow_{u, v}}{\leq \servicelimite_u * \instances_u~~~\forall u \in V}{\label{formal:instance_egress}}
\addConstraint{\textstyle\sum_{v \in V} \connections_{u,v}}{\leq \connectionlimit * \instances_v~~~\forall u \in V}{\label{formal:connection_egress}}
\addConstraint{\textstyle\sum_{u \in V} \connections_{u,v}}{\leq \connectionlimit * \instances_u~~~\forall v \in V}{\label{formal:connection_ingress}}
\addConstraint{N_v}{\leq \maxinstances~~\forall v \in V}{\label{formal:max_instancess}}
\end{argmini!}

\subsection{Throughput maximizing overlay paths}
\label{sec:solver:max_throughput}
Directly solving for a throughput maximizing path under a cost ceiling is non-trivial as we cannot use the linear reformulation of the cost objective. We can approximate a solution by solving for the minimum cost transfer plan at a range of many throughput goals. The result of this procedure is a Pareto frontier curve (as shown in \figref{fig:eval:solver:viz_tradeoff_curve}). A throughput maximizing solution can be extracted from this curve. The quality of approximate solution will depend on how many samples are used. A single AWS \texttt{c5.9xlarge} instance can evaluate 100 samples in under 20 seconds.

\section{Implementation of \sys{}}

We implemented \sys{} in Python 3.
\sys{}'s planner uses the proprietary Gurobi library to solve MILP instances (used in our evaluation), but the Coin-OR library can be used instead to avoid this dependency.
Our implementation currently supports the three major cloud providers: Amazon Web Services, Microsoft Azure, and Google Cloud Platform.

We use \texttt{m5.8xlarge} instances on AWS, as smaller VM sizes were subject to burstable networking performance, which we wished to avoid~\cite{aws_vm_egress_limit,aws_general_purpose_instances}.
For consistency, we used \texttt{Standard\_D32\_v5} instances on Microsoft Azure and \texttt{n2-standard-32} instances on Google Cloud.

A user initiates a transfer from their application with the \emph{\sys{} client}. The client provisions VMs in each region according to the transfer plan and runs the \emph{\sys{} gateway} program on each VM. The gateway is responsible for actually reading from source object stores, relaying data through overlay regions and writing to destination object stores.

While transfer time is dominated by network throughput, the time to spawn gateway VMs contributes to the transfer latency.
To minimize unnecessary bloat in VM images, we use compact OSes such as Bottlerocket~\cite{aws_bottlerocket} and package dependencies via Docker.

\sys{} assumes that objects are broken up into small \textit{chunks} of approximately equal size.
Applications can often do this without significant burden; for example, machine learning applications store data as \texttt{TFRecord}s, which are easy to split into small chunks.
This allows \sys{} to read and write data quickly from and to cloud object stores, by issuing many read/write operations in parallel to different chunks.

To mitigate the impact of straggler connections, \sys{} dynamically partitions data across TCP connections as they become ready to accept more data.
This is in contrast to tools like GridFTP~\cite{gridftp}, which assign data blocks to connections in a round-robin fashion.
The downside is that, for plans that use multiple overlay paths, the amount of data sent on each path may deviate from the targets computed at planning time, which could cause the actual cost of transferring data to deviate from the cost predicted by \sys{}'s planner.

To avoid overflowing buffers at relay regions, \sys{} uses hop-by-hop flow control to stop reading data from incoming TCP connections when a VM's queue of chunks reaches capacity.
Bufferbloat-type problems~\cite{bufferbloat} are not a concern for \sys{}, with regard to queued chunks, as we pipeline transfers to optimize for throughput instead of latency.
\section{Evaluation}
\label{sec:eval}

\begin{figure*}[t]
    \centering
    \subfloat[\centering AWS DataSync comparison]{{\includegraphics[width=0.33\textwidth]{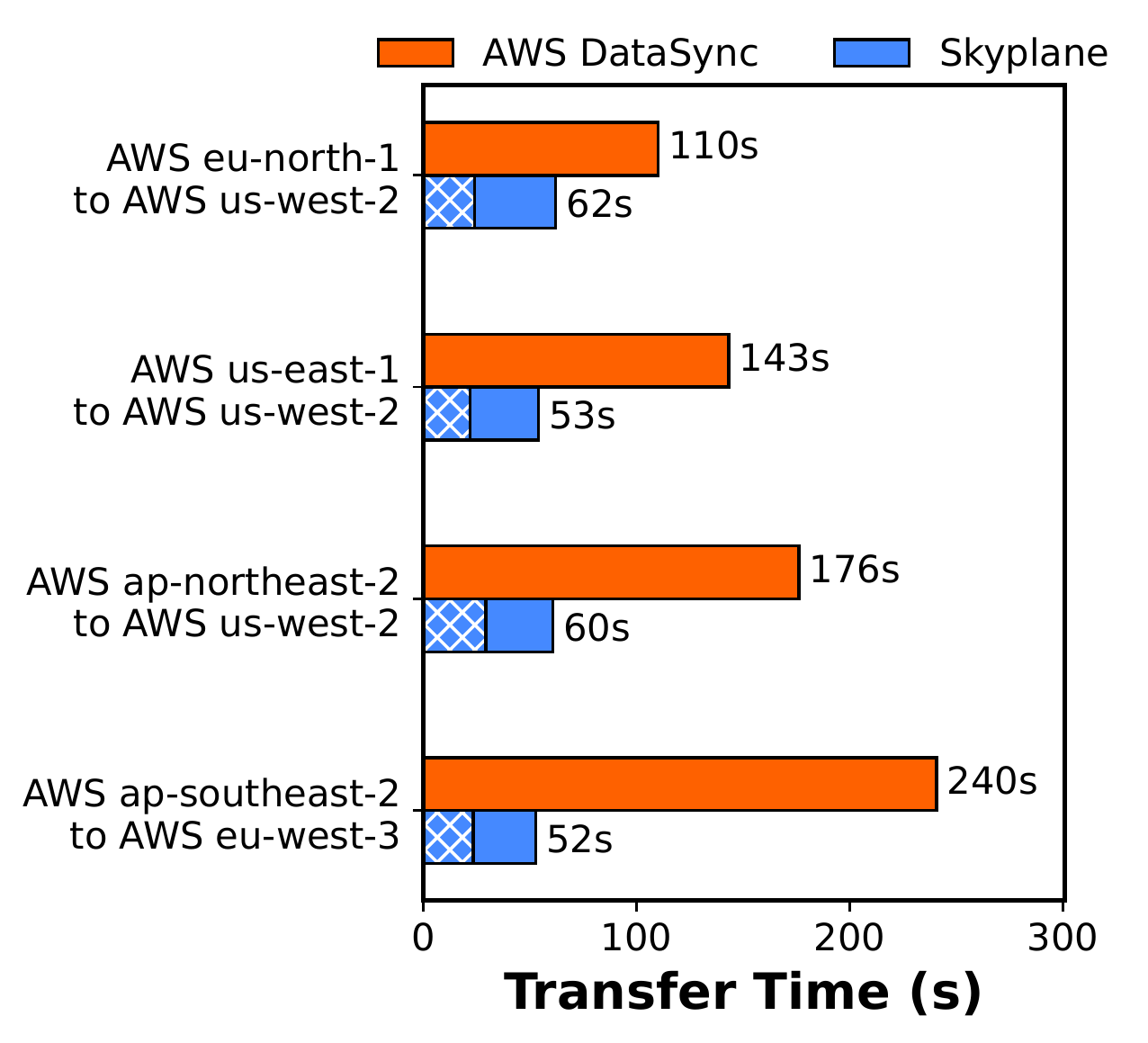}}}%
    \subfloat[\centering GCP Storage Transfer comparison]{{\includegraphics[width=0.35\textwidth]{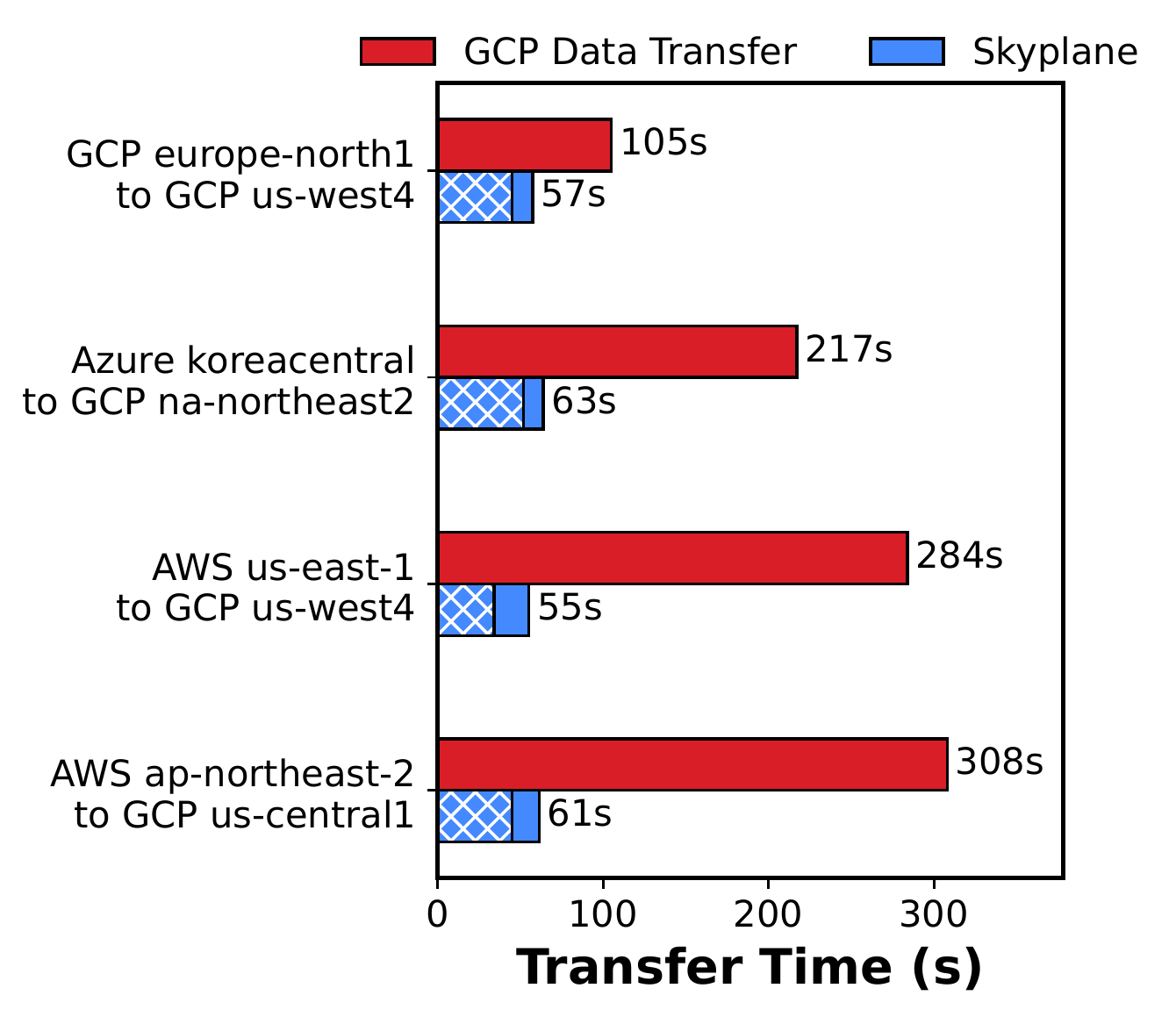} }}%
    \subfloat[\centering Azure AzCopy comparison]{{\includegraphics[width=0.33\textwidth]{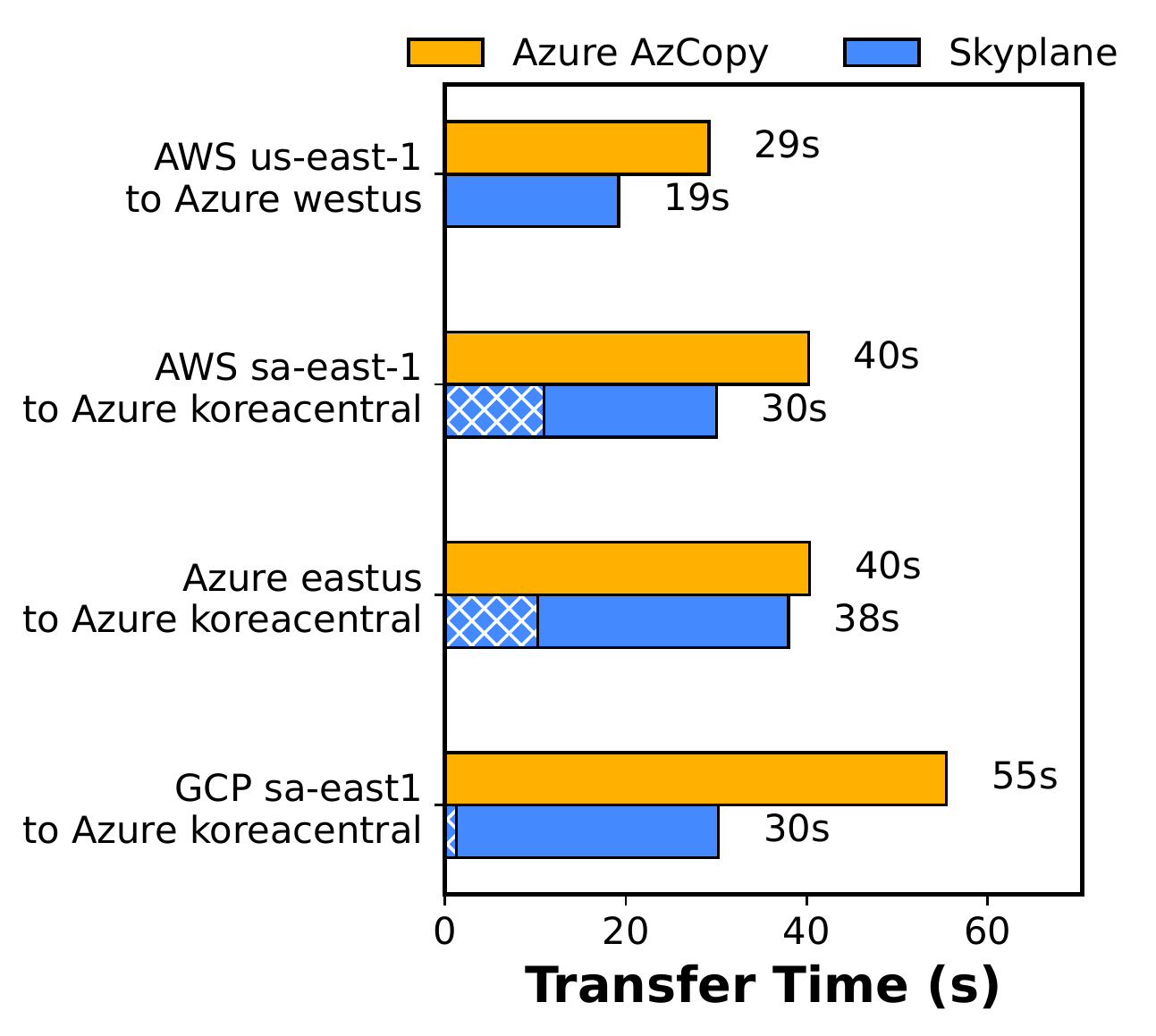}}}
    \caption{\textbf{Comparison to cloud transfer systems}: The thatch pattern in each bar represents the storage I/O overhead.
}
    \label{fig:cloud_transfer_comparision}
\end{figure*}

\begin{figure*}[t]
    \centering
    \includegraphics[width=\textwidth]{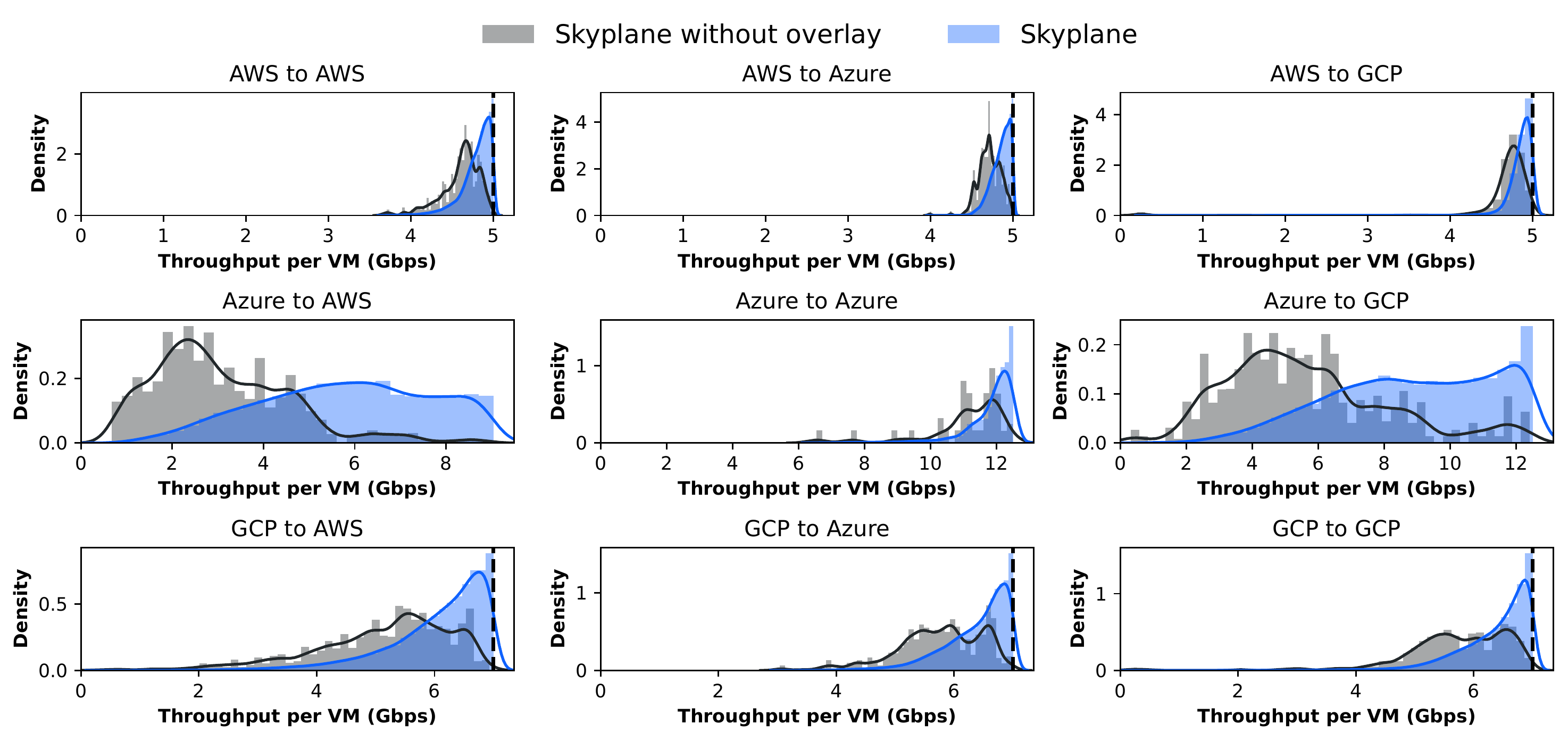}
    \caption{\textbf{Ablation of predicted overlays}: Overlay routes improve throughput per VM instance. We visualize the distribution of predicted throughput by the planner with all optimizations enabled (\sys{}) and with all optimizations except for overlay routing (\sys{} without overlay). The AWS and GCP egress limits are displayed with a dashed line.}
    \label{fig:eval:solver:overlay_ablation}
\end{figure*}

To evaluate \sys{}, we investigate transfer time and price.
We will sometimes use transfer throughput as a proxy for transfer time.
In our price calculations, we include both instance cost and egress cost.

\subsection{Experimental setup}
We evaluate \sys{} with 20 AWS regions, 24 Azure regions and 27 GCP regions.
For all experiments, we use public IP addresses attached to the VMs for transferring data.
In some cases, one can achieve better performance for intra-cloud overlay hops by using private IP addresses assigned to each VM.
For GCP this yields higher performance; for AWS and Azure it may yield higher performance, but requires peering virtual networks which incurs additional fees.

Furthermore, Azure and GCP allow one to select \emph{network tiers} to control whether data is transferred via the cloud provider's network or via the public Internet.
The \sys{} prototype utilizes external IPs over standard network tiers.
That said, \sys{} is not incompatible with optimizations like VPC peering or hot-potato routing tiers to reduce cost and improve performance which we leave to future work.
We use the CUBIC congestion control protocol in experiments.

\subsection{How much faster is \sys{} than existing data transfer solutions?}
\label{sec:eval:baselines}
Existing cloud providers offer data transfer tools such as AWS DataSync, GCP Storage Transfer, and Azure AzCopy for low-cost transfers of bulk data into their respective clouds.
These tools do not disclose what mechanisms they use to transfer data---for example, the number of VMs and TCP connections (if any) used for a transfer, or the QoS (if any) associated with the network traffic.
When evaluating \sys{}, we restrict \sys{} to use at most 8 VMs in each region.
This is conservative; for example, on equalizing \$/GB for some routes, \sys{} could provision \textit{up to 262 VMs} per region within DataSync's service fee. 
Moreover, while these services only support data transfer \emph{into} their respective clouds, \sys{} supports data transfer between every region pair.

We consider transferring the training and validation set for ImageNet~\cite{cloudtpu_example_resnet}.
We specifically use the \texttt{TFRecord}s as generated by Google as part of the Cloud TPU benchmark example~\cite{cloudtpu_example_resnet}.
We evaluate flows between regions within a single cloud (intra-provider) and between clouds (inter-provider).
We expected that data transfer within each cloud provider (e.g., between AWS's \texttt{us-east-1} and AWS's \texttt{us-west-1}) to perform well as they have full visibility into their networks and can utilize private interfaces with higher performance than over public API.
For example, Azure Blob Storage throttles per-object reads for third-party VMs\cite{azure_blob_throttle}.
Our experiments did observe this behavior.
However, \sys{} benefits from parallelizing the transfers.

We compare against AWS DataSync, GCP Storage Transfer and Azure AzCopy in \figref{fig:cloud_transfer_comparision}.
We evaluated \sys{} with a cost budget cap that is lower than the service fee for cloud transfer services in all our experiments.
For each source-destination pair, we additionally measured the time to transfer procedurally-generated data using \sys{}; this allows us to break out the overhead of reading and writing to cloud storage as a ``thatched'' region in each bar.
\sys{} significantly outperforms AWS DataSync and GCP Cloud Transfer in all configurations.
In certain cases, Azure AzCopy performs about as well as \sys{}.
We chose the \texttt{koreacentral} region because we expected the greatest improvements from the overlay in that region; however, storage overheads (the ``thatched'' regions of the bars), not networking overheads, dominated the runtime.
It is possible that AzCopy avoids the Azure Blob Storage I/O overhead that dominates \sys{}'s transfer time by leveraging Azure's \texttt{Copy Blob From URL} API call to download data directly into the servers running Azure Blob Storage~\cite{azcopy_copy_from_url}.

\subsection{How much faster are the overlay paths?}
\label{sec:eval:overlay}
The planner optimally explores the trade-off between improved throughput and cost for cloud data transfers.
We explore solving for the optimal transfer path between all pairs of clouds regions between all cloud providers.
We evaluated 22 AWS regions, 23 unrestricted Azure regions and 27 GCP regions which leads to 5,184 possible replication routes.
It would be too expensive to transfer a large amount of data along each path in order to measure the empirical achieved throughput; therefore we use the planner to generate a plan and compare the resulting plan with the direct path, both in terms of expected throughput and cost.
We compute predicted costs for transferring a 50 GB dataset between each possible source and destination.
We report the speedup relative to \sys{} with a direct connection between each set of instances.
Notice that the baseline is itself an ablation of \sys{} and it generally outperforms existing cloud transfer services to begin with (see \secref{sec:eval:baselines}).

\begin{figure}
    \centering
    \includegraphics[width=0.85\linewidth]{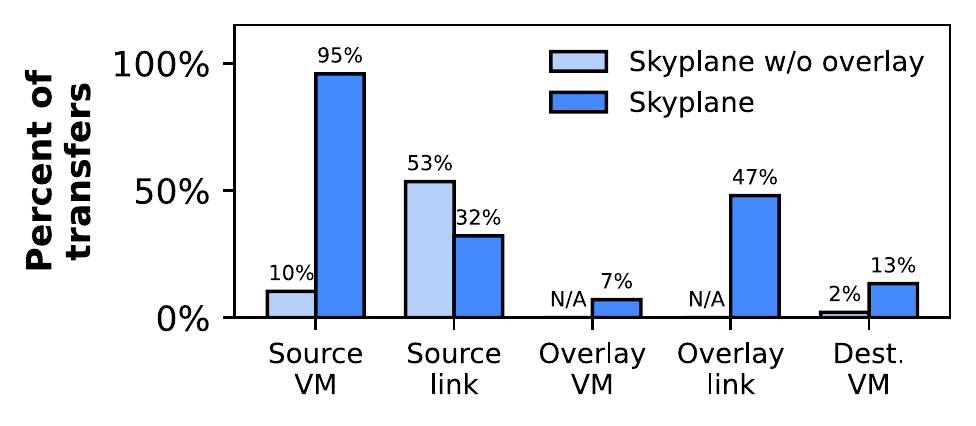}
    \caption{\textbf{Transfers bottlenecked at each location}: For transfers in Fig.~\ref{fig:eval:solver:overlay_ablation}, we visualize what percentage of transfers were bottlenecked at various locations. Enabling the overlay shifts bottlenecks from the network to the VM.} \label{fig:eval:solver:bottlenecks}
\end{figure}

The results are shown in \figref{fig:eval:solver:overlay_ablation}.
For each pair of source and destination clouds, we show distribution of predicted throughputs across region pairs, both with \sys{}'s planner restricted to the direct path and allowing \sys{}'s planner to use overlay paths.
The results show that \sys{}'s overlay routing meaningfully improves achievable throughput between cloud regions.
Note that transfers out of AWS cannot exceed 5 Gbps and transfers leaving GCP cannot exceed 7 Gbps due to these cloud providers' caps on egress bandwidth.

\subsection{Where are transfer bottlenecks?}

To understand how the overlay improves throughput, we characterize the fraction of transfers that are bottlenecked at each location. In \figref{fig:eval:solver:bottlenecks}, we visualize the percentage of transfers from \secref{sec:eval:overlay} that were bottlecked at a VM in the source region, the network link leaving the source region, a VMs in optional overlay regions, a network links leaving an overlay region, and a VM in the destination region. We consider a particular location to be a bottleneck if utilization is over $99\%$. Multiple locations may simultaneously be a bottleneck for one transfer.

For \sys{} with overlay routing disabled, the network link from the source to the destination region is the most common bottleneck for transfers. In a small set of cases, the source VM is a bottleneck for the transfer. Generally, the direct path is not fast enough to saturate the maximum egress bandwidth limit for a VM.
The overlay shifts source link bottlenecks by reducing the number of transfers bottlenecked by the source link by $32\%$. The bottleneck shifts to the source VM or in some cases a network link leaving an overlay region.

\begin{figure*}[t]%
    \centering
    \subfloat[TCP connections versus throughput]{\includegraphics[width=0.3\linewidth]{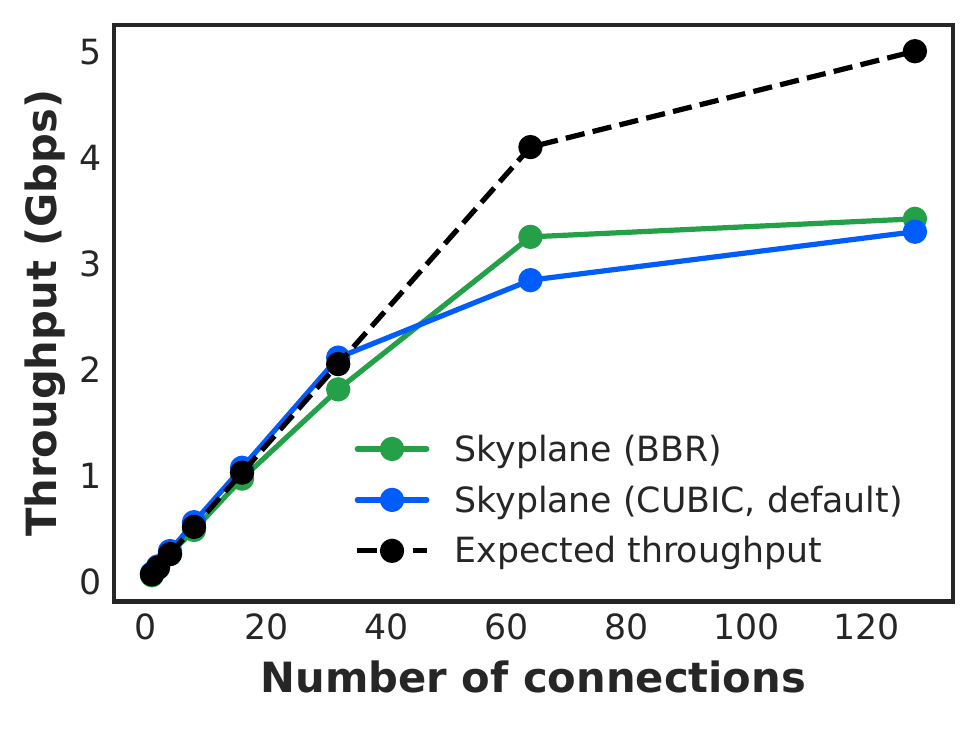}\label{fig:eval:system_ablation:tcp_connections}}\quad
    \subfloat[Number of gateway VMs versus throughput]{\includegraphics[width=0.3\linewidth]{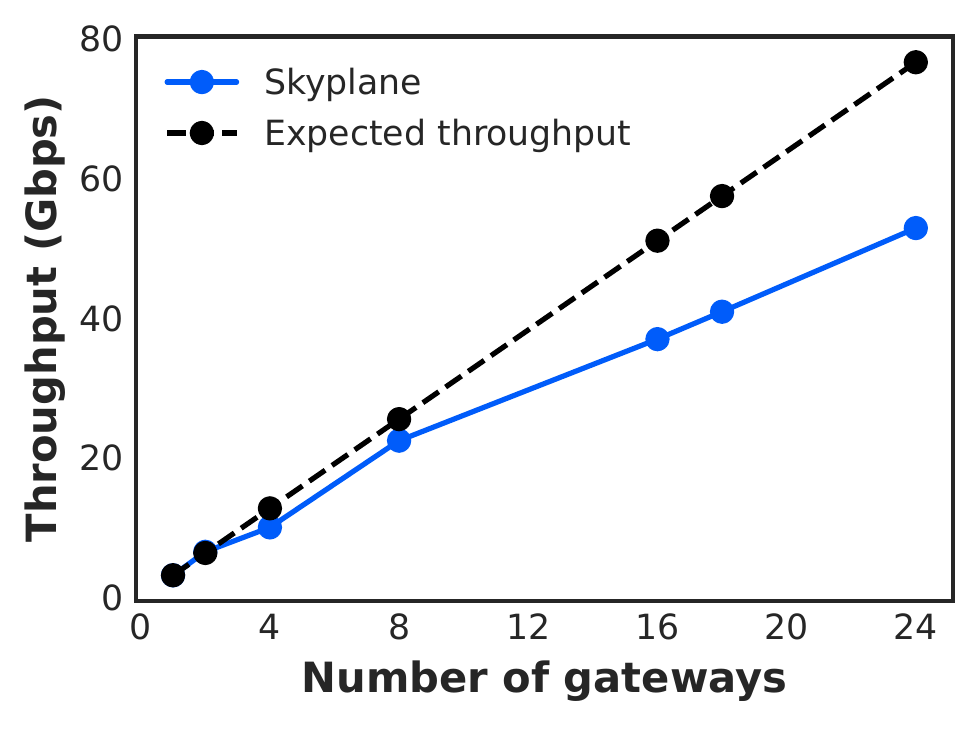}\label{fig:eval:system_ablation:parallel_vms}}\quad
    \subfloat[Predicted planner throughput versus cost\label{fig:eval:solver:viz_tradeoff_curve}]{ \includegraphics[width=0.3\linewidth]{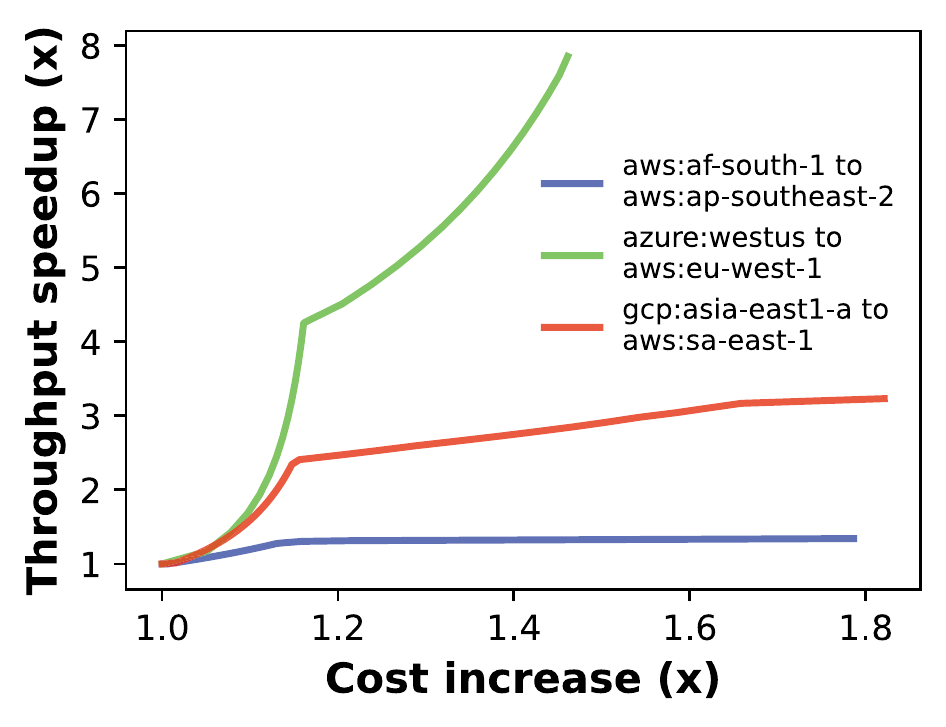}\label{fig:eval:system_ablation:throughput_vs_cost}}
    \caption{\textbf{\sys{} ablations}: We evaluate the impact of parallel TCP connections, parallel gateway VMs and overlay cost.}
    \label{fig:eval:system_ablation}
\end{figure*}

\begin{figure}
    \centering
    \includegraphics[width=\linewidth]{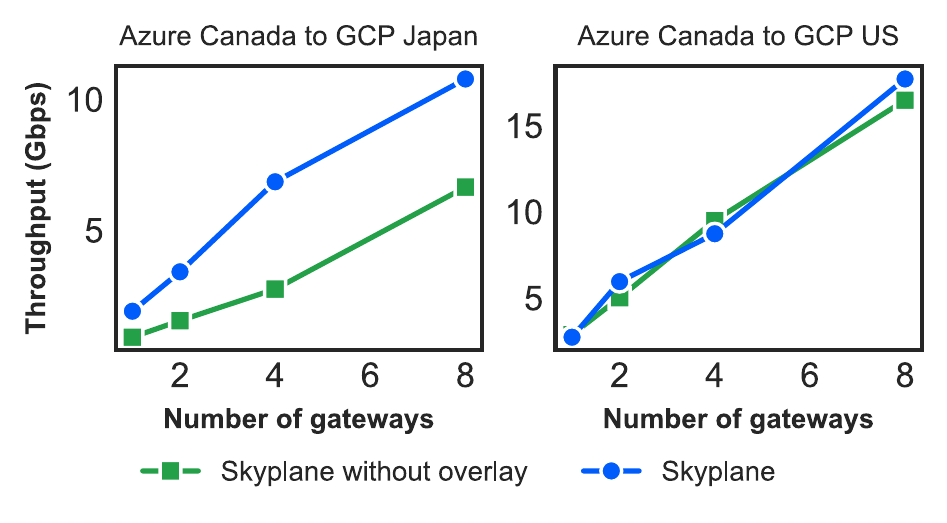}
    \caption{\textbf{Scaling VMs versus overlay}: In situations where the direct path is slow, the overlay is faster than simply scaling the number of VMs used alone.}
    \label{fig:ablation:vm_scalability}
\end{figure}

\subsection{\sys{} microbenchmarks}
\paragraph{Impact of parallel TCP connections}
\figref{fig:eval:system_ablation:tcp_connections} shows the impact of varying the number of parallel TCP connections used to transfer data between VMs.
For this experiment, the source VM was located in AWS \texttt{ap-northeast-1} and the destination VM was located in AWS \texttt{eu-central-1}.
\sys{} transfers 32 GB of synthetic, procedurally-generated data in these experiments to avoid incurring object store I/O overheads and thereby isolate network performance.
The black dashed line shows the expected throughput, assuming that bandwidth scales linearly with the number of parallel TCP connections up to AWS' 5 Gbps egress cap.
The blue line shows \sys{}'s achieved throughput, and the green line uses \sys{}'s achieved throughput using the BBR congestion control algorithm (used only this experiment).
For this experiment, the source VM was located in AWS \texttt{ap-northeast-1} and the destination VM was located in AWS \texttt{eu-central-1}.
\sys{}'s achieved throughput plateaus below the 5 Gbps egress cap, and 64 connections is enough to come close.

\vspace{-8pt}
\paragraph{Impact of parallel VMs}
\figref{fig:eval:system_ablation:parallel_vms} shows the impact of using multiple VMs in each region to achieve higher aggregate throughput.
The black dashed line shows the expected throughput, assuming that bandwidth scales linearly with the number of VMs.
Although \sys{}'s performance is significantly less than the expected throughput for a large number of gateways, the graph shows that using parallel VMs is an effective way for \sys{} to scale its aggregate bandwidth.
Additionally, using parallel VMs is a particularly valuable tool in the context of inter-cloud transfers, as \sys{} can use multiple VMs in one cloud provider to circumvent the egress limit.
For example, for an overlay hop from an AWS region to an Azure region, one may allocate many instances in AWS but few in Azure, to account for AWS' egress cap.

\vspace{-8pt}
\paragraph{Trade-off between cost and throughput}
\label{sec:eval:tradeoff_cost_throughput}
\figref{fig:eval:system_ablation:throughput_vs_cost} shows the impact on overlay path throughput as the price budget is varied.
We adjusted the cost budget afforded to the planner (x-axis), and plot the throughput predicted by the planner for the output plan (y-axis).
We show three routes where the overlay benefits are considerable (Azure \texttt{westus} to AWS \texttt{eu-west-1}), good (GCP \texttt{asia-east1-a} to AWS \texttt{sa-east-1}) and minimal (AWS \texttt{af-south-1} to AWS \texttt{ap-southeast-2}).
As the cost budget increases, \sys{} uses increasingly complex overlay topologies, adding new overlay paths as the instance limit (1 VM, in this case) is saturated in each region.
Each elbow in the plot (e.g. 1.2$\times$ for the Azure to AWS route) represents a point where \sys{} adds a new overlay route via a faster but more costly region.
At some point, the planner cannot increase throughput further as the overlay network is saturated.

\vspace{-8pt}
\paragraph{Is it better to use VMs to form overlay paths or parallelize the direct path?}
Given a limited number of VMs (\secref{s:multiple_instances}), a natural question is whether it is better to use those VMs to form overlay paths or to parallelize the direct path.
In \figref{fig:ablation:vm_scalability}, we evaluate \sys{} with and without the overlay enabled for various numbers of VMs in the context of an inter-continental transfer and an intra-continental transfer.
For the inter-continental transfer, using the VMs with overlays enabled provides a 2.08$\times$ geomean speedup compared to using those VMs to parallelize the direct path.
However, for the intra-continental transfer, there is little benefit to using VMs in overlay paths (1.03$\times$ geomean speedup).

\begin{table}[t]
\centering
\caption{\textbf{Comparison with academic baselines}: \sys{} outperforms RON's path selection heuristic implemented in \sys{}~\cite{ron}.}
\label{tab:academic_eval_comparison}
\resizebox{\linewidth}{!}{%
\begin{tabular}{@{}lccc@{}}
\toprule
\multicolumn{1}{c}{\textbf{Method}} & \multicolumn{1}{c}{\textbf{Time}} & \multicolumn{1}{c}{\textbf{Throughput}} & \multicolumn{1}{c}{\textbf{Cost}} \\ \midrule
GCT GridFTP~\cite{gridftp,gridftp_gct} (1 VM) &  133s & 1.03 Gbps & \$1.40\\
\cellcolor{Gray}\sys{} (1 VM, direct) & \cellcolor{Gray}  \textbf{73s} & \cellcolor{Gray} \textbf{1.71 Gbps} & \cellcolor{Gray} \$1.40 \\ \midrule
\sys{} w/ RON routes (4 VMs)~\cite{ron} &  21s &  6.02 Gbps &  \$2.27 \\
\sys{} (cost optimized, 4 VMs) &   32s &   3.88 Gbps &  \$1.56 \\
\cellcolor{Gray}\sys{} (throughput optimized, 4 VMs) & \cellcolor{Gray} \textbf{16s} & \cellcolor{Gray}  \textbf{8.07 Gbps} & \cellcolor{Gray}  \$1.59 \\ \bottomrule
\end{tabular}%
}
\end{table}

\subsection{Comparison against academic baselines}
In \autoref{tab:academic_eval_comparison}, we compare \sys{} with RON~\cite{ron} and the community-maintained fork~\cite{gridftp_gct} of GridFTP~\cite{gridftp} for a 16 GB data transfer from Azure \texttt{East US} to AWS \texttt{ap-northeast-1}.
To isolate network throughput from I/O overheads, we benchmark the transfers without object stores (VM to VM only).

We use the open-source GCT fork of GridFTP~\cite{gridftp_gct}.
Although GCT GridFTP theoretically supports striped transfers across multiple machines, we were unable to find a supported non-commercial implementation.
To make a fair comparison, we run both GCT GridFTP and Skyplane with a single VM per region. 
\sys{} is $1.6\times$ faster than GCT GridFTP.

We implement RON's path selection heuristic in \sys{} to compare overlays between RON and \sys{}. 
Our results show that \sys{} has better cost and throughput than RON.
\sys{} with routes from RON's path selection heuristic achieves $3.5\times$ higher throughput than \sys{} with a single VM but at $62\%$ cost overhead.
\sys{}'s planner instead finds overlay paths with up to $4.7\times$ higher throughput than the direct path within a $14\%$ cost overhead.
\section{Related Work}

\sys{} builds on the overlay network literature~\cite{ron, cronets, akamai}.
As discussed in \secref{sec:intro}, \sys{} adapts classical overlays to the cloud setting, accounting for the price of network bandwidth and leveraging the elasticity of cloud resources.
CRONets~\cite{cronets} briefly discusses cost, but focuses on comparing cloud-based options to private leased lines.
Unlike \sys{}, it does not discuss how to manage the cost of cloud resources.
Lai et al.~\cite{to_relay_or_not} find relay regions improve throughput in AWS when utilizing a single TCP connection but find the 2 Gbps instance NIC limit from their chosen instance class limits the benefit of overlay paths.
CloudCast~\cite{cloudcast_vmware} examines the use of triangle overlays in the cloud to reduce network latency while \sys{} examines throughput.

Several existing efforts~\cite{mccat, jetstream, clon} aim to optimize bulk data transfers by reducing the amount of data transferred.
Such techniques are complementary to \sys{}; one can first apply these techniques to reduce the amount of data to transfer, and then apply \sys{}'s techniques to transfer that reduced data efficiently.
Unlike \sys{}, these works do not use cost when selecting the network path to use for a transfer.

Another line of research aims to improve bulk data transfers by improving resource management.
GridFTP~\cite{gridftp} is a tool for wide-area transfers that techniques such as using multiple machines and TCP connections.
GridFTP sends all data over the direct path and does not utilize overlays.
Khanna et al.~\cite{globus_overlay} explore application of network overlays to GridFTP but do not consider elasticity and egress price in the cloud.
Other solutions, like PSockets~\cite{psockets}, also use parallel TCP connections for high bandwidth.
Pied Piper~\cite{piedpiper} also explored how cloud resource elasticity could be used to improve cloud data transfers, but utilize a different mechanism than \sys{}.

There have been decades of improvements and optimizations at the transport layer to make TCP perform better in large-BDP settings within TCP itself~\cite{rfc2488, rfc7323, cubic, bbr}, while others concern operating system support for TCP~\cite{userspacetcp,upcalls,fbufs}.
Improvements to TCP are complementary to \sys{}.
CodedBulk~\cite{codedbulk} uses network coding to complete bulk-transfer multicast jobs quickly~\cite{codedbulk}.
Another set of research~\cite{amoeba, bulkgeosdn, deadlinegrid} investigates how to schedule urgent and non-urgent bulk transfers to meet a transfer's deadline.
None of these techniques consider the cost of transferring data in the cloud.

Traffic engineering (TE) systems, like Google's B4~\cite{b4, b4after} and BwE~\cite{bwe} and Microsoft's SWAN~\cite{swan}, Cascara~\cite{cascara}, and BlastShield~\cite{blastshield}, are used internally by cloud providers to navigate the cost-performance trade-off in their wide-area networks.
The precise nature of the trade-off differs from \sys{} in two ways.
First, TE systems consider costs in terms of the \emph{bandwidth} provisioned (e.g., the cost of installing long-distance cables~\cite{b4}, or the 95th percentile bandwidth for peering links~\cite{cascara}).
In contrast, \sys{} considers cost from the perspective of a cloud customer, where the cost depends on the volume and not bandwidth of data transferred.
Second, TE systems like Cascara~\cite{cascara} assume a static topology and aim to reallocate bandwidth to save cost, with a global view of a single provider's network. \sys{} optimizes a single user's transfer, with the ability to use overlay regions in multiple cloud providers' networks.

\sys{} has similarities to Content Delivery Networks (CDNs)~\cite{akamai}, most notably in that both make use of overlay networks.
However, \sys{}'s focus is different from CDNs.
CDNs focus on caching objects near users, in order to provide low network latency.
In contrast, \sys{} focuses on transferring large amounts of data quickly, with a focus on achieving high bandwidth rather than low network latency such as in workloads like ML training and database replication.
CDNs are more suitable for workloads where popular objects need to be replicated to many regions so that geo-distributed users can access them with low network latency.

One application of bulk transfers is VM migration~\cite{snowflock,elephants_dance,livemigration_vm,supercloud} that balance VM downtime and bandwidth consumed when transferring VMs. Supercloud~\cite{supercloud} uses a network of vSwitches in an overlay that maintains TCP connections upon migration, not to provide high bandwidth at low cost.

Some existing research efforts and commercial products focus on bulk transfer jobs that are not time-critical.
For example, Laoutaris et al.~\cite{delaytolerantbulktransfers} propose techniques to reduce the cost of transferring data for delay tolerant applications.

Cloud providers provide services for bulk transfer, such as AWS Snowball~\cite{snowball}, Azure Data Box~\cite{azure_data_box}, and GCP Transfer Appliance~\cite{gcp_transfer_appliance}, that have users ship their data via physical drives via the postal service.
For sufficiently large transfers, these services may allow data to be transferred into the cloud datacenter more quickly than using the Internet.
\section{Conclusion}

This paper explores how to efficiently transfer data between cloud regions using cloud-aware overlay networks.
Our key observation is that principles from overlay networks can be applied to the cloud setting to identify high-quality network paths that lead to fast transfer times.
However, adapting principles from overlay networks to the cloud setting requires consideration of cloud resource pricing, most notably the egress fees associated with network bandwidth.
\sys{} manages the trade-off between performance and cost when performing bulk data transfer.
It works by accepting a user- or application-provided constraint on performance and solving a mixed integer linear program (MILP) to obtain the optimal data transfer plan.
\sys{} can reduce the time to transfer data by up to $5.0\times$ at minimal additional cost.

\section*{Acknowledgments}
We thank the anonymous reviewers and our shepherd, Rachee Singh, for their helpful feedback.
We also thank Asim Biswal, Jason Ding, Daniel Kang, Vincent Liu, Xuting Liu, and Anton Zabreyko.
This work is supported by NSF CISE Expeditions Award CCF-1730628, NSF GRFP Award DGE-1752814, and gifts from Amazon, Astronomer, Google, IBM, Intel, Lacework, Microsoft, Nexla, Samsung SDS, and VMWare.

\clearpage
\bibliographystyle{plain}
\bibliography{reference}

\end{document}